\documentclass{article}
\usepackage{natbib}
\usepackage{authblk}
\usepackage{rotating}
\begin{document}

\newcommand\aap{A\&A}                % Astronomy and Astrophysics
\let\astap=\aap                          % alternative shortcut
\newcommand\aapr{A\&ARv}             % Astronomy and Astrophysics Review (the)
\newcommand\aaps{A\&AS}              % Astronomy and Astrophysics Supplement Series
\newcommand\actaa{Acta Astron.}      % Acta Astronomica
\newcommand\afz{Afz}                 % Astrofizika
\newcommand\aj{AJ}                   % Astronomical Journal (the)
\newcommand\ao{Appl. Opt.}           % Applied Optics
\let\applopt=\ao                         % alternative shortcut
\newcommand\aplett{Astrophys.~Lett.} % Astrophysics Letters
\newcommand\apj{ApJ}                 % Astrophysical Journal
\newcommand\apjl{ApJ}                % Astrophysical Journal, Letters
\let\apjlett=\apjl                       % alternative shortcut
\newcommand\apjs{ApJS}               % Astrophysical Journal, Supplement
\let\apjsupp=\apjs                       % alternative shortcut
% The following journal does not appear to exist! Disabled.
%\newcommand\apspr{Astrophys.~Space~Phys.~Res.} % Astrophysics Space Physics Research
\newcommand\apss{Ap\&SS}             % Astrophysics and Space Science
\newcommand\araa{ARA\&A}             % Annual Review of Astronomy and Astrophysics
\newcommand\arep{Astron. Rep.}       % Astronomy Reports
\newcommand\aspc{ASP Conf. Ser.}     % ASP Conference Series
\newcommand\azh{Azh}                 % Astronomicheskii Zhurnal
\newcommand\baas{BAAS}               % Bulletin of the American Astronomical Society
\newcommand\bac{Bull. Astron. Inst. Czechoslovakia} % Bulletin of the Astronomical Institutes of Czechoslovakia 
\newcommand\bain{Bull. Astron. Inst. Netherlands} % Bulletin Astronomical Institute of the Netherlands
\newcommand\caa{Chinese Astron. Astrophys.} % Chinese Astronomy and Astrophysics
\newcommand\cjaa{Chinese J.~Astron. Astrophys.} % Chinese Journal of Astronomy and Astrophysics
\newcommand\fcp{Fundamentals Cosmic Phys.}  % Fundamentals of Cosmic Physics
\newcommand\gca{Geochimica Cosmochimica Acta}   % Geochimica Cosmochimica Acta
\newcommand\grl{Geophys. Res. Lett.} % Geophysics Research Letters
\newcommand\iaucirc{IAU~Circ.}       % IAU Cirulars
\newcommand\icarus{Icarus}           % Icarus
\newcommand\japa{J.~Astrophys. Astron.} % Journal of Astrophysics and Astronomy
\newcommand\jcap{J.~Cosmology Astropart. Phys.} % Journal of Cosmology and Astroparticle Physics
\newcommand\jcp{J.~Chem.~Phys.}      % Journal of Chemical Physics
\newcommand\jgr{J.~Geophys.~Res.}    % Journal of Geophysics Research
\newcommand\jqsrt{J.~Quant. Spectrosc. Radiative Transfer} % Journal of Quantitiative Spectroscopy and Radiative Transfer
\newcommand\jrasc{J.~R.~Astron. Soc. Canada} % Journal of the RAS of Canada
\newcommand\memras{Mem.~RAS}         % Memoirs of the RAS
\newcommand\memsai{Mem. Soc. Astron. Italiana} % Memoire della Societa Astronomica Italiana
\newcommand\mnassa{MNASSA}           % Monthly Notes of the Astronomical Society of Southern Africa
\newcommand\mnras{MNRAS}             % Monthly Notices of the Royal Astronomical Society
\newcommand\na{New~Astron.}          % New Astronomy
\newcommand\nar{New~Astron.~Rev.}    % New Astronomy Review
\newcommand\nat{Nature}              % Nature
\newcommand\nphysa{Nuclear Phys.~A}  % Nuclear Physics A
\newcommand\pra{Phys. Rev.~A}        % Physical Review A: General Physics
\newcommand\prb{Phys. Rev.~B}        % Physical Review B: Solid State
\newcommand\prc{Phys. Rev.~C}        % Physical Review C
\newcommand\prd{Phys. Rev.~D}        % Physical Review D
\newcommand\pre{Phys. Rev.~E}        % Physical Review E
\newcommand\prl{Phys. Rev.~Lett.}    % Physical Review Letters
\newcommand\pasa{Publ. Astron. Soc. Australia}  % Publications of the Astronomical Society of Australia
\newcommand\pasp{PASP}               % Publications of the Astronomical Society of the Pacific
\newcommand\pasj{PASJ}               % Publications of the Astronomical Society of Japan
\newcommand\physrep{Phys.~Rep.}      % Physics Reports
\newcommand\physscr{Phys.~Scr.}      % Physica Scripta
\newcommand\planss{Planet. Space~Sci.} % Planetary Space Science
\newcommand\procspie{Proc.~SPIE}     % Proceedings of the Society of Photo-Optical Instrumentation Engineers
\newcommand\rmxaa{Rev. Mex. Astron. Astrofis.} % Revista Mexicana de Astronomia y Astrofisica
\newcommand\qjras{QJRAS}             % Quarterly Journal of the RAS
\newcommand\sci{Science}             % Science
\newcommand\skytel{Sky \& Telesc.}   % Sky and Telescope
\newcommand\solphys{Sol.~Phys.}      % Solar Physics
\newcommand\sovast{Soviet~Ast.}      % Soviet Astronomy (aka Astronomy Reports)
\newcommand\ssr{Space Sci. Rev.}     % Space Science Reviews
\newcommand\zap{Z.~Astrophys.}       % Zeitschrift fuer Astrophysik

\title{X-ray luminosity versus orbital period of AM CVn systems}
\author[1]{Teja Begari}
\author[2]{Thomas J. Maccarone}
\affil[1]{Independent Scientist, Hyderabad, India, begariteja@gmail.com}
\affil[2]{Department of Physics \& Astronomy, Texas Tech University, Lubbock TX 79409}
\date{}
\maketitle

\begin{abstract}
   AM CVn systems are a rare type of cataclysmic variable star consisting of a white dwarf accreting material from a low-mass, hydrogen-poor donor star. These helium-rich systems usually have orbital periods that are less than 65 minutes and are predicted to be sources of gravitational waves.  We have analyzed the catalogued X-ray data from the Chandra, XMM-Newton, and The Neil Gehrels Swift Observatory (hereafter referred to as 'Swift') to investigate the relationship between X-ray luminosity and the orbital period of AM CVn systems. We find that the high accretion-rate systems which are likely to have optically thick boundary layers are sub-luminous in X-rays relative to theoretical model predictions for the boundary layer luminosity, while the longer orbital period, lower bolometric luminosity systems match fairly well to the model predictions, with the exception of an overluminous system which has already been suggested to show magnetic accretion. 
\end{abstract}

\section{Introduction}
AM CVn stars are binary systems that have very short orbital periods that range from 5 to 65 minutes. These systems consist of white dwarfs accreting material from Roche lobe-filling companion star that usually is a lower-mass white dwarf star, but occasionally is a helium star.  These systems are expected to be strong sources of gravitational waves \citep{2004MNRAS.349..181N}. In this paper, we discuss the relationship between X-ray luminosity and orbital period for a sample of 28 AM CVn systems. We use the data from XMM, Chandra, and Swift observatories for collecting the X-ray flux and obtaining Orbital Period values from the literature. Studying the relationship between X-ray luminosity and orbital period can provide insights into the accretion process in AM CVn systems and techniques for searching for more of them.

\section{Data and analysis}
We collected the X-ray flux from Chandra, XMM-Newton, and Swift for all AM CVn systems listed in Table 1. Using the Chandra Source Catalog (CSC), we collected the flux from 0.5-7.0 keV from Release 2.0 \citep{2010ApJS..189...37E} For XMM-Newton, using 4XMM-DR11, we collected the flux from 0.2-12.0 keV from \citet{2020A&A...641A.136W}. For Swift, using the 2SXPS Swift X-ray telescope point source catalogue , we collected the flux from 0.3-10.0 keV from \citet{2022A&A...657A.138T}.  We collected the distances from Gaia parallax measurements corrected for the Gaia zero-point offset, and which used Bayseian analysis to convert the measured parallaxes into inferred distances \citep{2021AJ....161..147B}. Using the flux we calculated the X-ray luminosity for each system. We also obtained the orbital periods of these systems from \citet{2018A&A...620A.141R}. For the newly discovered AM CVn system, TIC 378898110, we collected the orbital period, X-ray flux and the distance from \citet{2023arXiv231101255G} In Table 1, we see that a few systems have periods estimated using superhumps.  Superhump periods are typically within a few percent of the real periods, which is acceptable for the purposes of comparing luminosities with orbital periods.  Using the uncertainty in the distances, we calculated 16\% and 84\% values for the luminosity. Using these values we made the error bars.  

Our primary analysis involved plotting the X-ray luminosity versus the orbital period for the sample of AM CVn systems. Figure 1 shows a comparison between the observational data and a model prediction for X-ray luminosity from \citet{2012A&A...537A.104V}, 

\begin{equation}
  L  = \frac{GM_a}{2R_a}\frac{48}{5}\frac{G^{2/3}}{2c^5}\frac{M_a (9 \pi)^2 10^{-6}M_\odot R_\odot^3}{\left(M_a + \frac{9 \pi 10^ {-3} \sqrt{\frac{M_\odot R_\odot ^ 3}{2G}}}{P_{orb}}\right)^ {1/3}} \frac{(2 \pi)^{8/3}}{P_{orb}^{14/3}}
\end{equation}
where $M_a$ is the mass of the accretor, $P_{orb}$ is the orbital period, and $R_a$ is the radius of the accretor over the observed data.  For plotting the predicted X-ray luminosity, we substituted the value of $M_a$ as $0.8 M_\odot$ from the work by \citet{2021ApJ...923..125W} and that the luminosity generated in the boundary layer is 1/2 of the total accretion power, and all of the boundary layer luminosity goes into X-rays and the evolution of ultra-compact binaries by \citet{2004MNRAS.349..181N,2012A&A...537A.104V}

We conduct a Spearman's rank correlation test for the sources that have the orbital period greater than 30 minutes, yielding a correlation coefficient of -0.48235 with a two-tailed p-value of 0.05846. This suggests a marginally significant association between the two variables. If we exclude the system SDSSJ0804+1616 and re-run the test, we obtained rs = -0.575, with a two-tailed p-value of 0.02494, again marginally significant

Our findings also indicate that, at short orbital periods, the accretion rates are high enough to keep the system in a state at which the system is constantly accreting material from its companion star at a very high rate, where the boundary layer is optically thick. In this regime, the X-rays produced at the white dwarf surface are thermalized into UV photons, leading to suppression of the X-ray emission, as seen in transient outbursts in dwarf novae \citep{2003MNRAS.345...49W}. 

Two systems stand out as overluminous relative to the trend of sources at similar orbital periods.  One is V407~Vul, the shortest period object, which is a direct impact accretor\citep{2002MNRAS.331L...7M}, which may lead to a higher fraction of its accretion power coming out in X-rays.  The other is SDSS~J0804+1616, which shows a strong magnetic field for the accretor, which may drive its orbital evolution to be faster than that due to gravitational radiation\citep{2023arXiv230212318M}.

\section{Conclusion}
In conclusion, we observe that AM CVn systems with short orbital periods have X-ray luminosities below those from literature model predictions, but which, in hindsight, should have been anticipated given the expectation that bright systems will have optically thick boundary layers. We find a clear anti-correlation between X-ray luminosity and orbital period for the longer orbital period systems, in agreement with the theoretical expectations for the accretion process in these systems.

\begin{figure}
    \includegraphics[width =0.7 \textwidth]{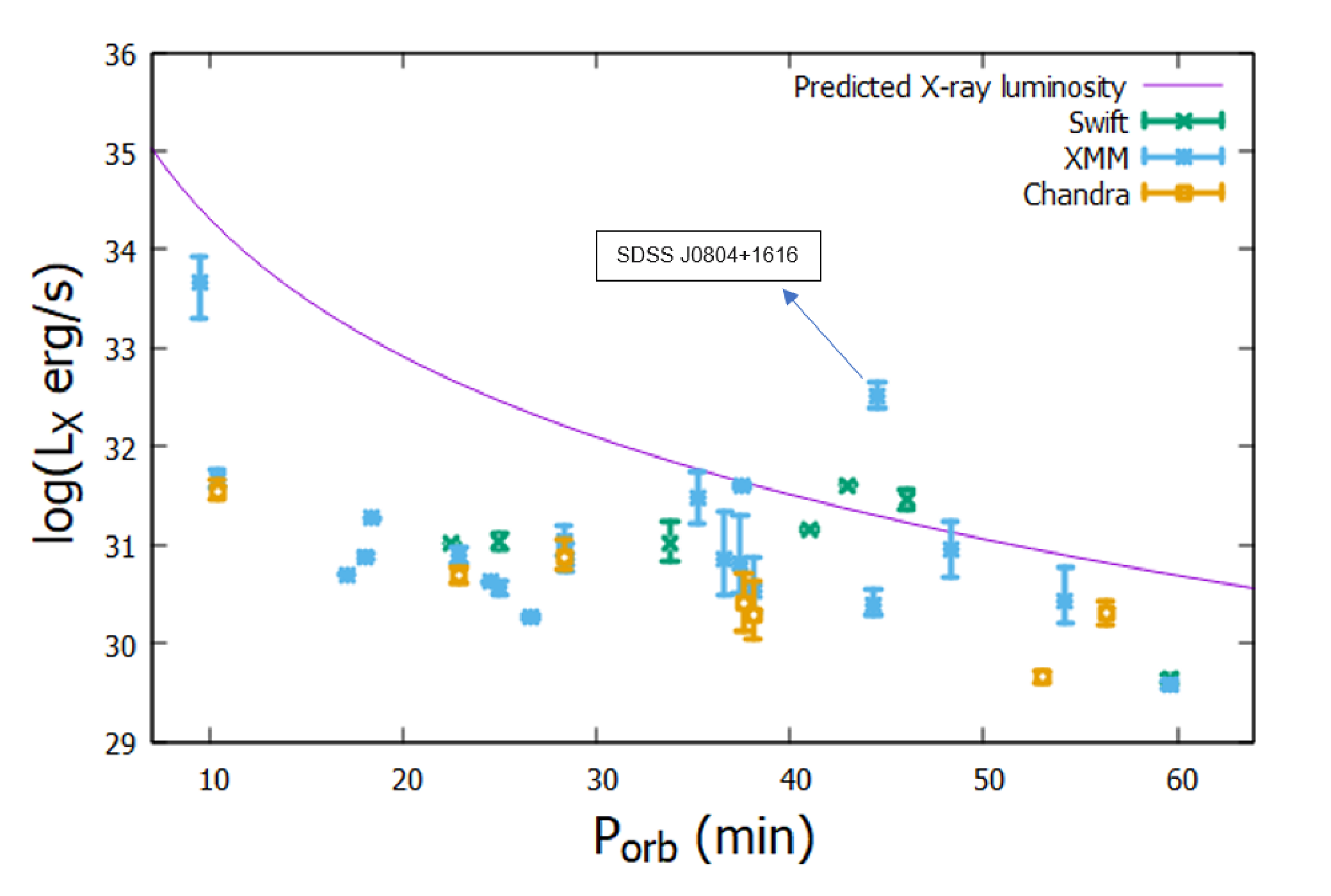} 
    \caption{X-ray luminosity versus orbital period of selected AM CVn systems from Table 1.}
\end{figure}

\begin{sidewaystable}[!ht]
    \centering
    \begin{tabular}{|l|l|l|l|l|l|l|l|}
    \hline
        ~ & ~ & Log(X-Ray luminosity) in erg/sec & ~ & ~ & Distance (in parasecs) & ~ & ~ \\ \hline
        Star System & P\_orb in mins  & XMM & Chandra & Swift & Best estimate & B\_pc(84\%) & b\_pc(16\%) \\ \hline
        V407 Vul & 9.5 & 33.66 & ~ & ~ & 4813.76  & 6584.68& 3214.56 \\ \hline
        ESCet & 10.4 & 31.67 & 31.55 & ~ & 1786.57 & 2013.34 & 1612.83 \\ \hline
        AM Cvn & 17.1 & 30.7 & ~ & ~ & 300.02 & 303.06 & 297.24 \\ \hline
        SDSS J1908+3940 & 18.1 & 30.88 & ~ & ~ & 968.33 & 1004.87 & 935.96 \\ \hline
        HP lib & 18.4 & 31.28 & ~ & ~ & 277.91 & 280.4 & 275.38 \\ \hline
        TIC 378898110 & 20.5 & ~ & ~ & 31.02 & 309.3 & 311.1 & 307.5 \\ \hline
        CX361 & 22.9 & 30.9 & 30.69 & ~ & 963.45 & 1058.7 & 878.92 \\ \hline
        CR Boo & 24.5 & 30.63 & ~ & ~ & 349.55 & 353.67 & 344.79 \\ \hline
        KL Dra & 25 & 30.56 & ~ & 31.04 & 907.92 & 985.06 & 829.11 \\ \hline
        V803 Cen & 26.6 & 30.27 & ~ & ~ & 284.08 & 290.12 & 279.26 \\ \hline
        YZ LMi & 28.3 & 31.04 & 30.89 & ~ & 815.72 & 994.31 & 694.03 \\ \hline
        CP Eri & 28.4 & 30.86 & ~ & ~ & 725.62 & 874.06 & 623.87 \\ \hline
        V406 Hya & 33.8 & ~ & ~ & 31.02 & 753.68 & 966.01 & 616.76 \\ \hline
        SDSS J1730+5545 & 35.2 & 31.49 & ~ & ~ & 1317.93 & 1771.03 & 977.98 \\ \hline
        V558 Vir & 36.6(sh) & 30.86 & ~ & ~ & 1548.05 & 2692.72 & 1015.78 \\ \hline
        SDSS J1240-0159 & 37.4 & 30.81 & ~ & ~ & 764.3 & 1358.88 & 544.51 \\ \hline
        NSV1440 & 37.5(sh) & 31.6 & ~ & ~ & 1861.73 & 1917.21 & 1814.65 \\ \hline
        SDSS J1721+2733 & 38.1 & 30.54 & 30.29 & ~ & 674.02 & 997.94 & 510.86 \\ \hline
        ASASSN-14mv & 41(sh) & ~ & ~ & 31.15 & 247.06 & 253.23 & 240.59 \\ \hline
        ASASSN-14ei & 43(sh) & ~ & ~ & 31.61 & 256.89 & 259.87 & 253.71 \\ \hline
        SDSS J1525+3600 & 44.3 & 30.4 & ~ & ~ & 538.6 & 639.85 & 469.63 \\ \hline
        SDSS J0804+1616 & 44.5 & 32.51 & ~ & ~ & 998.97 & 1184.42 & 865.22 \\ \hline
        SDSS J1411+4812 & 46 & ~ & ~ & 31.46 & 452.01 & 502.2 & 397.95 \\ \hline
        SDSS J0902+3819 & 48.3 & 30.96 & ~ & ~ & 709 & 976.22 & 512.44 \\ \hline
        SDSS J1208+3550 & 53 & ~ & 29.66 & ~ & 210.7 & 223.64 & 198.89 \\ \hline
        SDSS J1642+1934 & 54.2 & 30.42 & ~ & ~ & 554.84 & 824.24 & 432.7 \\ \hline
        SDSS J1552+3201 & 56.3 & ~ & 30.31 & ~ & 422.52 & 481.38 & 369.72 \\ \hline
        SDSS J1137+4054 & 59.6 & 29.58 & ~ & 29.64 & 209.06 & 218.5 & 199.36 \\ \hline
    \end{tabular}
    \caption{ 
    Note: b\_pc and B\_pc refer to the lower and upper limit in the uncertainty in the the distances, respectively. Sh indicates that the orbital period was derived from superhumps. }
\end{sidewaystable}

\bibliographystyle{aasjournal}
\bibliography{References}

\end{document}